\documentclass[10pt,journal,trans]{IEEEtran}
\usepackage{graphicx, wrapfig, subcaption, setspace, booktabs}
\usepackage{float}
\usepackage{enumerate}
\usepackage{scrextend}
\usepackage{chngpage}
\usepackage{lipsum}
\usepackage{calc}
\usepackage{lipsum}
\usepackage{multicol}
\usepackage[font=small,labelfont=bf]{caption}
\usepackage{graphicx}
\usepackage{amsmath}
\usepackage{amssymb}
\usepackage{lipsum}
\usepackage{tabularx}
\usepackage{amsmath}
\usepackage{algorithm}
\usepackage{array}
\usepackage{xcolor}
\usepackage[noend]{algpseudocode}
\usepackage{authblk}

\makeatletter
\def\BState{\State\hskip-\ALG@thistlm}
\makeatother
\ifCLASSOPTIONcompsoc
  \usepackage[nocompress]{cite}
\else
  \usepackage{cite}
\fi
\begin{document}

\title{Critical Impact of Social Networks Infodemic on Defeating Coronavirus COVID-19 Pandemic: Twitter-Based Study and Research Directions}

\author{
\IEEEauthorblockN{A. Mourad\IEEEauthorrefmark{1}, A. Srour\IEEEauthorrefmark{1}, H. Harmanani \IEEEauthorrefmark{1}, C. Jenainati\IEEEauthorrefmark{2}, M. Arafeh\IEEEauthorrefmark{1}\\
}
\IEEEauthorblockA{\IEEEauthorrefmark{1}Department of Computer Science \& Mathematics\\}
\IEEEauthorblockA{\IEEEauthorrefmark{2}Department of English\\}
\IEEEauthorblockA{Lebanese American University, Lebanon}

\thanks{ \textcircled c 2020 IEEE. Personal use of this material is permitted. Permission from IEEE must be obtained for all other uses, in any current or future media, including reprinting/republishing this material for advertising or promotional purposes, creating new collective works, for resale or redistribution to servers or lists, or reuse of any copyrighted component of this work in other works.}

}

%\author{A. Mourad,~\IEEEmembership{Senior Member,~IEEE,}
%        A.~Srour,
%        H.~Harmanani,~\IEEEmembership{Senior Member,~IEEE,}
 %       ~C.~Jenainati,
%        and M.~Arafeh% <-this % stops a space
%\IEEEcompsocitemizethanks{
%\IEEEcompsocthanksitem A. Mourad, A. Srour, H. Harmanani, and A. Arafeh are with the Department of Computer Science and Mathematics, Lebanese American University, Lebanon.\protect\\
%E-mail: see azzam.mourad@lau.edu.lb
%\IEEEcompsocthanksitem {C. Jenainati is with the Department of English, Lebanese American University, Lebanon.}% <-this % stops a space
%}
%}%

%\author{A. Mourad, A. Srour, H. Harmanani, C. Jenainati and M. Arafeh}%
%\affil{Lebanese American University, Lebanon}

\IEEEtitleabstractindextext{%
\begin{abstract}
News creation and consumption has been changing since the advent of social media. An estimated $2.95$ billion people in 2019 used social media worldwide. The widespread of the Coronavirus COVID-19 resulted with a tsunami of social media. Most platforms were used to transmit relevant news, guidelines and precautions to people. According WHO, uncontrolled conspiracy theories and propaganda are spreading faster than the COVID-19 pandemic itself, creating an {\em infodemic} and thus causing psychological panic, misleading medical advises, and economic disruption. Accordingly, discussions have been initiated with the objective of moderating all COVID-19's communications, except those initiated from trusted sources such as the WHO and authorized governmental entities. This paper presents a large-scale study based on data mined from {\tt Twitter}. Extensive analysis has been performed on approximately $1$ million COVID-19 related tweets collected over a period of two months. Furthermore, the profiles of $288,000$ users were analyzed including unique users' profiles, meta-data and tweets' context. The study noted various interesting conclusions including the critical impact of the (1) exploitation of the COVID-19 crisis to redirect readers to irrelevant topics and (2) widespread of unauthentic medical precautions and information. Further data analysis revealed the importance of using social networks in a global pandemic crisis by relying on credible users with variety of occupations, content developers and influencers in specific fields. In this context, several insights and findings have been provided while elaborating computing and non-computing implications and research directions for potential solutions and social networks management strategies during crisis periods.

\end{abstract}
}

\maketitle

\IEEEdisplaynontitleabstractindextext
\IEEEpeerreviewmaketitle

\section{Introduction}

\IEEEPARstart{T}{h}e worldwide spread of the COVID-19 infectious disease resulted with a pandemic that has threatened millions of lives. Social media has been playing a major role in fighting the virus and its impact through a multitude of measures including the continuous transmission of local and global updates about the pandemic as well as issuing warnings and and guidelines for dealing with the pandemic and its aftermath. According to Statista \cite{statista}, an estimated 2.95 billion people in 2019 used social media worldwide. The number is projected to increase to 3.43 billion in 2023. One remarkable statistic is around the continually changing demographic of new consumers and the increase in social media penetration reach. For example, while in 2018 the Pew Research Centre \cite{pew} reported that “most Americans continue to get news on social media, even though they may have concerns about its accuracy”. Numerous surveys have been undertaken to capture the online behavior of news consumers worldwide, and the trend seems to be that social media platforms are highly influential when it comes to acquiring news stories, for the majority of people. In a large-scale study conducted in 2019 by Ofcom \cite{Ofcom}, the UK government’s regular for the communications services that are used by the public, it was shown that “Half of the adults in the UK now use social media to keep up with the latest news". Furthermore, governments and major centers for disease control, including the World Health Organization (WHO) and the Centers for Disease Control and Prevention (CDC), are relying on social networks as a mean for managing the evolving pandemic by regularly disseminating guidance and updates and by providing emergency responses. 

The dark side of social media was exhibited in a tsunami of fake and unreliable news that ranged from selling fake cures to using the social media as a platform to launch cyberattacks on critical information systems. This led the {\em United Nations} to warn against a proliferation of false information about the virus and the emergence of the COVID-19 {\em infodemic}, according to WHO Director-General Tedros Adhanom Ghebreyesus at the Munich Security Conference on Feb 15, 2020 \cite{infodemic}. Moreover, various researchers and news outlets \cite{c1, c2, c3, c5, c6, c7, c8, c9} tackled the rising {\em infodemic} issue and presented real-life case studies detailing actual examples that impeded people from acting appropriately during the {\em infodemic}. For example, malicious users have used social media platforms such as {\em Facebook}, {\em Twitter}, {\em Instagram}, {\em Youtube} and {\em WhatsApp} in order to spread panic and confusion through deliberate overabundance of misleading information and rumors. A notable false claim that 5G damages the immune system and consequently causes the COVID-19 outbreak went viral and resulted with various burning of cell towers in Europe \cite{5g}. Other conspiracy theories spread rumors regarding the source and cure for COVID-19 at a time when people needed to focus during the outbreak on how to do the right thing in order to control the disease and mitigate its impact (e.g. virus does not infect children, virus dies in temperature above 27 degrees, combination of diet offering cures and immunity for the virus, cure discovery). Cyberattacks also flourished during the outbreak \cite{c6}. Videos, photos and posts in different languages exploited the COVID-19 context in order to redirect the general public to shady websites and inadvertently install spyware. Some cybersecurity firms claimed that 3-8\% of the newly registered COVID-19 related sites are suspicious, while others phishing messages about potential cures lead to the installation of malware.  

The {\em infodemic} resulted with organizations, governments and business leaders exercising excessive pressure on the social media platforms in order to curtail the flood of fake news and viral misinformation. This became a priority in order to ensure that people who are in lockdown would receive the appropriate information in order to do the appropriate thing, control the disease and mitigate its impact. Although social networks platforms have plans for mitigating and banning harmful content, it is apparent that they themselves were not well prepared and needed an emergency plan in order to respond to COVID-19 {\em infodemic}. In fact, most platforms are now filtering and banning users who are identified as sources of verified misinformation. However, this has led at times to more misinformation and at other times to accounts being unfairly removed as social media platforms do not have the capability of dealing with a huge amount of misleading and unverified data. The main focus was on advertisement and offering personalized services to both industries and people while analyzing human behavior and preference. Moreover, in most cases they used machine learning and artificial intelligence tools resulting in a lot of false positives. Although there are many various {\em intelligent} approaches in the literature tackling the identification of credible content in social media, the topic was not of high priority for the research and industrial communities. In fact, there was no justification for investing in this research direction.

This paper aims to address the aforementioned problems while tackling the evolving challenges using a large dataset that was extracted from twitter targeting COVID-19. The study uses a data analytics approach based on tweets meta-data, text and context, as well as users meta-data and profiles. The paper explores extensively one million COVID-19-related tweets that were collected over a period of two months and belonging to 288K users. The analysis of the unique users' profiles, meta-data and context of the tweets allowed us to deduce various important findings and insights while providing guidance for potential solutions. To the best of our knowledge, except Li et al. \cite{c12} who characterized the propagation of situational information in social media during COVID-19, no computing-related work has yet empirically addressed the positive or negative impact of social networks {\em infodemic} during the COVID-19 crisis. Accordingly, this paper contributes to the field by highlighting based on empirical analysis several findings and directions to a research field to become of great importance in the near future.

In what follows, we provide a summary of our findings. Please note the following terms usage in the paper: a {\em Tweet} refers to a unique tweet excluding the retweets, {\em Interactions} refer to the total number of retweets and favorites per unique tweet, and {\em Reach} refers to total number of followers of the user who initiated the unique tweet and reflects the number of tweeters that may potentially see and interact with it. The initial results indicate that around 16.1\% of the tweets (i.e. 160K Tweets, 2.1M Interactions and 5.6B Reach) are exploiting COVID-19 contexts for advertisement, redirecting users to out of scope topics or even maliciously misleading the community. A further ontology-based analysis on the context and users' meta-data confirms that only 3.5\% of the unique users initiating the tweets have medical profile while 2.8\% are virus specialists. Accordingly, at least of 93.7\% of the COVID-related tweets (i.e. 800K Tweets, 17M Interactions and 30B reach) may be transmitting misleading or unverified medical information. On the other hand and in order to highlight the importance of non-medical users in spreading important information in such situation, a deeper analysis was performed to identify unique users with key specialties. Results reaffirmed our initial findings and show that users with context-relevant occupations such as doctor, writer, reporter, journalist, editor and governor do not even constitute 1\% of the total reach count (i.e. 300M out of 37B).  Accordingly, these insights illustrated the need to identify relevant influencers in specific contexts and seek their help in order to disseminate verified and reliable information. Finally, it is important to note that the infodemic that is impacting social media including Facebook, Instagram, Snapchat, etc. is by order of magnitude bigger \cite{c4}. 

The contributions of this work are three folds:
\begin{itemize}
\item Empirical study providing quantitative measurement of the critical impact of social networks {\em infodemic} during COVID-19 pandemic. To the best of our knowledge, no computing-related work has yet addressed and studied through experiments either the positive or the negative impact of social networks on defeating COVID-19.
\item Mixed ontology-based data analytics methodology and real-life experiments with inferred insights targeting both user profiles and tweet contexts for (1) detecting tweets exploiting COVID-19 for spreading misleading information and (2) identifying the source of tweets per user speciality and occupation for measuring the credibility and reliability of the disseminated COVID-19-related information.
\item Elaboration of both computing and non-computing findings, implications, social networks management strategies and research directions supported with thorough literature review for a field to become of great importance in the near future.
\end{itemize}

The remainder of this paper is organized as follows. In Section \ref{Methodology}, we illustrate the study's research methodology while in Section \ref{exploit} we provide an analysis of the impact of misleading twitter contexts. Section \ref{credibility} provides empirical analysis of the impact of COVID-19 related posts per user specialty and occupation.  Section \ref{directions} details our research findings and directions while Section \ref{conclusion} concludes with comments.  

\section{Methodology and Data Processing}
\label{Methodology}

\begin{figure}
\centering
\includegraphics [scale=0.63]{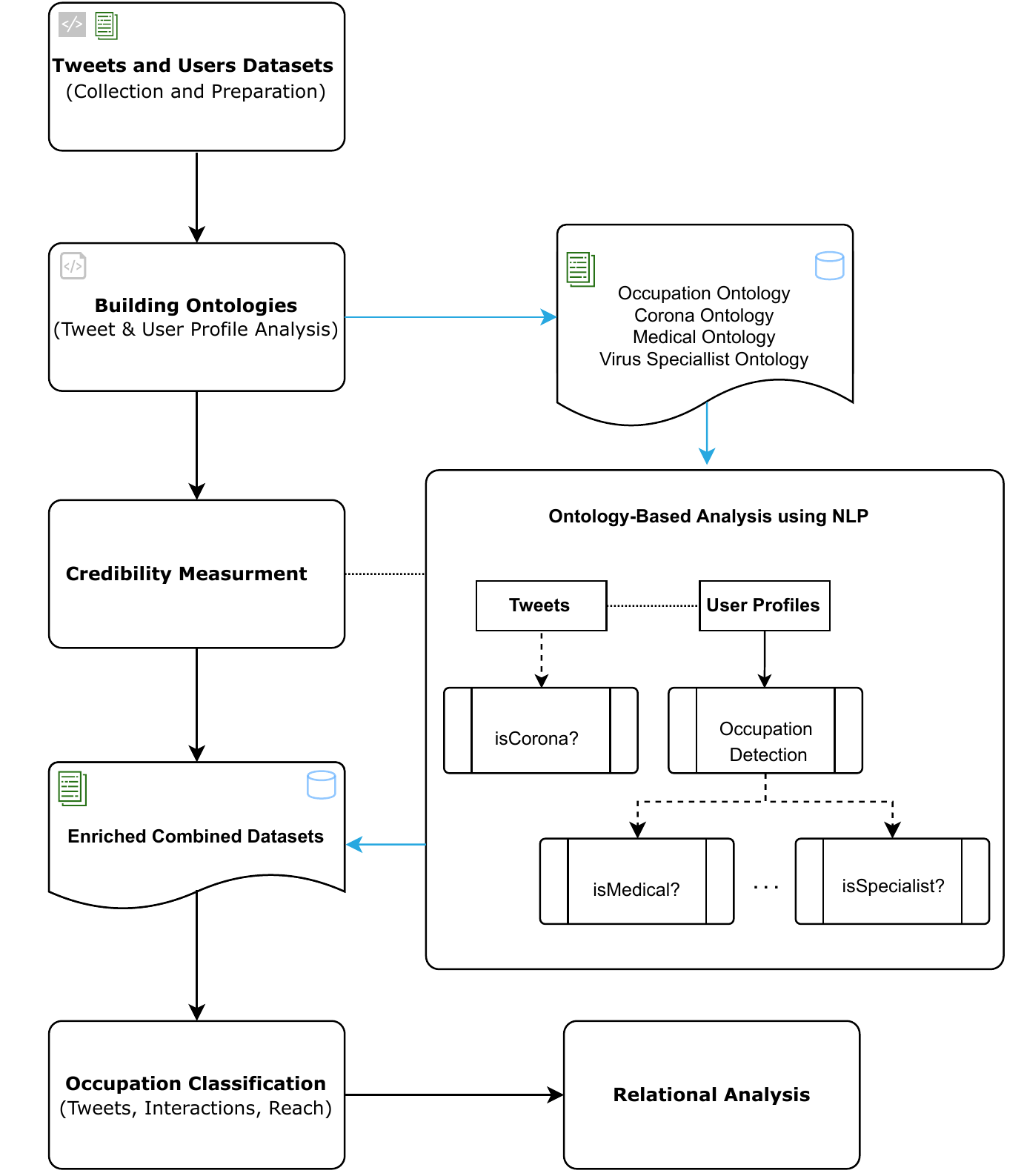}
\caption{Methodology Overview}
  \label{fig:fig0}
\end{figure}

The adopted methodology, illustrated in Figure \ref{fig:fig0}, starts by choosing the pertinent topic and selecting the top used hashtags. A search query that forms the basis of the data collection scripts was next built and the keywords were selected. The system systematically fetched approximately a million tweets from {\tt Twitter} along with their corresponding users' profiles. A descriptive analysis report was then generated by aggregating the collected records. In order to gain deeper insight into the collected data, we developed five different ontology relationships. The ontology rules allowed the system to analyze the content and consequently build the targeted aggregations. Natural Language Processing (NLP) techniques were used in order to classify the tweets and the users based on the above analysis. Finally, the results were aggregated and inferences were made based on users' occupations. 

In the sequel, we provide an ordered and detailed description of the methodology presented in Figure \ref{fig:fig0} including the proposed approaches and elaborated solutions within each of the system modules:

\begin{itemize}
\item A crawler Python script was implemented using a {\tt tweepy} \cite{c71} for collecting one million public tweets that include the ”corona” or ”covid” terms. The keywords were selected based on the the top used hashtag strings in the event from the result of {\tt Twitter Advanced Search Query} output. Once the data is fetched, a list of unique users who initiated the tweets was extracted and {\tt Twitter REST API} \cite{c72} access tokens were used in order to fetch the public profiles and perform the aggregations and analysis.

\item A set of ontologies and lexicons was built based on special keyword list to classify tweets into corona or non-corona related ones and infer insights from tweets and user profile datasets. In this regard, the ontologies were used as a base for the NLP entity extractor to classify each tweet based on its content regardless of the hashtags. Similarly, they were also used to classify users that have medical and speciality backgrounds. The following are the five built ontologies: {\em Corona Top Used Hashtags Ontology, Corona Social Media Context Ontology for Tweets, Occupation Ontology for Grouping Users Based on their Biographic Information, Medical Occupation Ontology for Users and Virus Specialty Occupation Ontology for Users}.

\item The ontology-based processing scripts were next built in order to extract entities from tweets as well as from the user biography fields. Accordingly, we inferred credibility measurement using NLP analysis. The distributed scripts simultaneously processed tweets and user records in order to tag record with a final value (i.e. {\em isCorona, isMedicalProfile, isSpecialtyProfile, isCoronaHashtag,} and the list of detected occupations).

\item  The dataset was next decorated for advanced filtering and analysis queries by merging the aggregated data into one enriched dataset, which was augmented with the following attributes: {\em Unique Tweet ID}, {\em Hashtag Counts per Tweet, Favorite Counts per Tweet, Retweet Counts per Tweet, Mention Counts per Tweet, Interactions (favorite and retweet) Counts per Tweet, Total Reach Count (number of followers per user per unique tweet), Unique User ID, Claimed Locations per User, Occupations per User (extracted from the user profile biography field), isCorona-Related (a boolean expression), isMedicalProfile-Related (a Boolean expression) and isSpecialtyProfile-Related (a boolean expression)}.

\item An occupation classification was next performed in order to better understand the effect of the tweets that were initiated by users with different roles and specialities. Within this context, we counted each user's unique tweets per occupation group (e.g. journalists), calculated the total Interactions per tweet, and calculated the total Reach counts caused by the mentioned group of users.

\item An Analysis of the correlation across users/groups that have medical profiles as well as a specialization in the study of viruses or infectious diseases was performed. Both profile types share similar entities and keywords, and thus we attempted to highlight and studied the impact of users with a virus specialization profiles rather than those with a general medical background by sub-categorizing users with medical profiles.
\end{itemize}

\section{Impact Analysis of Tweets Exploiting COVID-19 Context}
\label{exploit}
\begin{figure}
\centering
\includegraphics [scale=0.55]{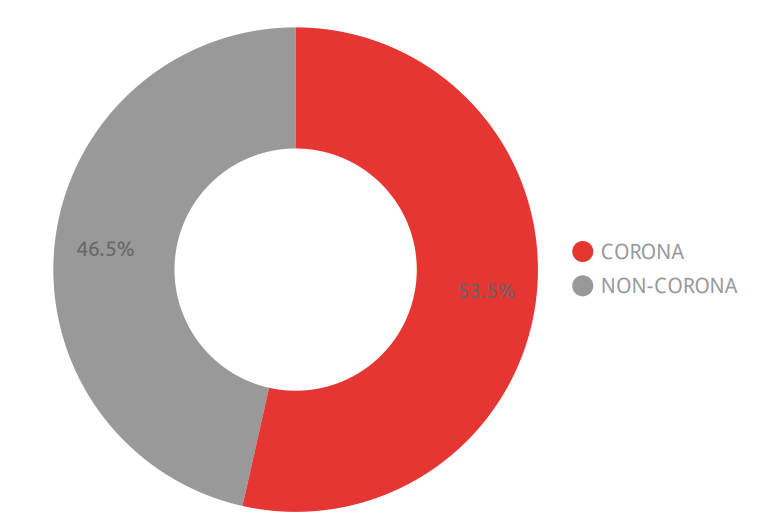}
\caption{CORONA vs NON-CORONA Related Hashtags}
  \label{fig:fig1}
\end{figure}

In this section, we present the main findings and discuss the insights and the results of the analysis based on the predefined framework approach and KPIs. As we processed around 109.3K hashtags from the one million random unique tweets, it was important to classify each hashtag according to it is direct relationship to the COVID family of hashtags. For instance, regardless of the context of the tweet, a hashtag that matches or partially contains {\em \#COVID} or {\em \#CORONA} is classified as {\em CORONA} since it is explicitly related to the Corona virus, while other hashtags like {\em \#China}, {\em \#US} or {\em \#Italy} are classified as {\em NON-CORONA} since they are not directly related to the Corona virus. Figure \ref{fig:fig1} shows the comparison between the occurrences of the two classes ({\em CORONA} and {\em NON-CORONA}) in the tweets. It can be noted that 53.5\% of the tweets (582.9K Tweets) represent tweets that are explicitly related to the Corona virus based on the hashtag, while 46.5\% (506.2K Tweets) represent tweets that are not explicitly related to the Corona virus and belong to the {\em NON-CORONA} classification based on the hashtags. It should be noted that since some tweets contain hashtags from both classes, the total number of the classified classes does not reflect the number of unique tweets but rather the count of tweets. This explains the fact that the number of tweets per each class does not add up to one million (same applies to Figure \ref{fig:fig3}).

\begin{figure}
\centering
\includegraphics[scale=0.58]{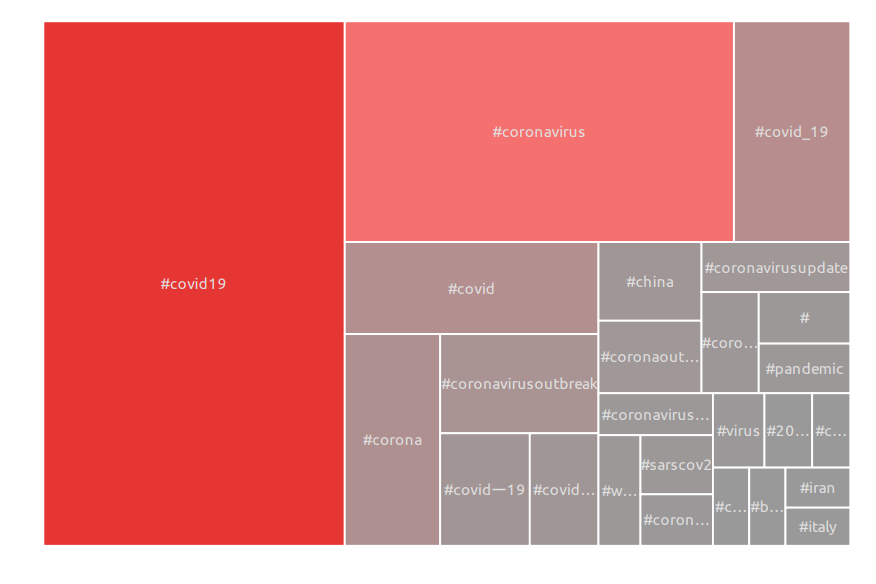}
\caption{Top Used Hashtags in Different Dialogs}
  \label{fig:fig2}
\end{figure}

\begin{figure}
\centering
\includegraphics[scale=0.7]{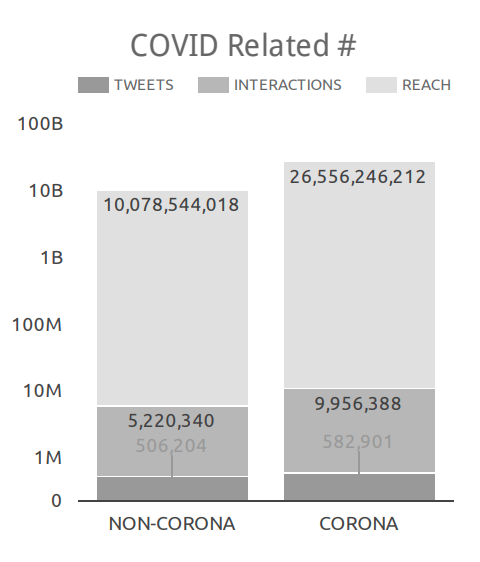}
\caption{Tweets, Interactions \& Reach Counts of COVID Related \& Non-Related Hashtags}
  \label{fig:fig3}
\end{figure}

Figure \ref{fig:fig2} displays the top used hashtags from the 109.3K ones sorted by the total count of occurrences in all tweets. The {\tt TreeMap} visualization chart has three dimensions to display. The position (from left to right), the box size (bigger to smaller), and the color opacity(100\% to 1\%). All dimensions are displayed based on the number of total occurrences of each hashtag in the entire tweets dataset. It should be also mentioned that the displayed hashtags have different dialogs and formats. For example, {\em Covid19, COVID19, and covid19} were counted as separate hashtags in order to measure the different usage for later text analysis searches.

Figure \ref{fig:fig3} shows the total Interactions and  Reach counts of each class of hashtags ({\em CORONA} and {\em NON-CORONA}) using a stacked column chart. It is interesting to notice that the number of Interactions and Reach level covered by the COVID hashtags on just a small set of users compared to the actual twitter size. The number of Interactions reflects the total Interactions (i.e. retweets and favorites) of all the unique tweets where the classified hashtags were used. The total Reach displays the possible Reach counts of the mentioned unique tweets based on their users' followers count. Again, both Reach and Interactions summations of the two classes do not sum up to the total Reach and Interactions specified in the header. We notice that the total number of Reach counts of the two classes is 36.6B out of 36.7B (a difference of 86,000,000 possible Reach), which indicates that more than 80\% of the users performed the discovered Unique Tweets.

\begin{figure}
\centering
\includegraphics[scale=0.6]{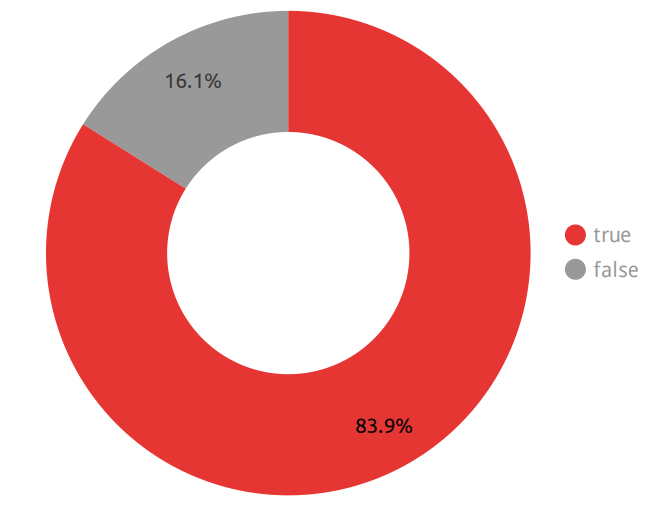}
\caption{Tweets Within and Diverting Out of COVID Context}
  \label{fig:fig4}
\end{figure}

\begin{figure}
\centering
\includegraphics[scale=0.7]{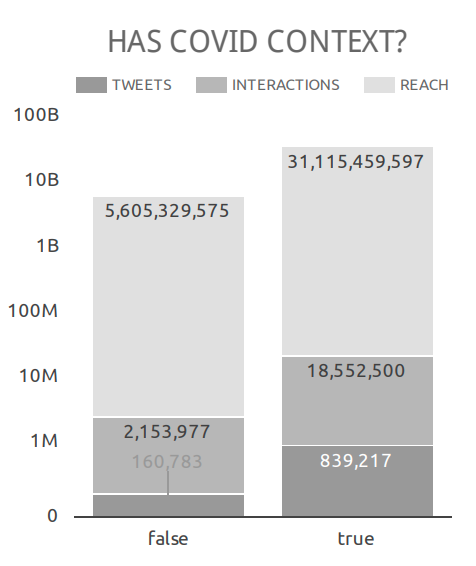}
\caption{Tweets, Interactions and Reach Counts Within and Diverting Out of COVID Context}
  \label{fig:fig5}
\end{figure}

Furthermore, an ontology-based classification of the contexts was performed in order to understand the meaning of the tweets. The ontology is built from COVID related dictionary for identifying the tweets diverting from the context to different topics. Figure \ref{fig:fig4} shows that 16.1\% of the tweets (i.e. 160.1K Unique Tweets) were not related to the COVID situation at all, while 83.9\% (839.2K Unique Tweets) were related based on their content. Some of the non-related ones were using the trend hashtags to advertise for products and other topics, and others were malicious intended to mislead the trend into different subjects. Figure \ref{fig:fig5} shows the total Interaction and Reach counts of each tweet in each classified category. In addition to the details mentioned in the description of Figure \ref{fig:fig4}, it is important to highlight the large effect of the 16.1\% tweets in terms of Interactions and Reach counts, which recorded around 2M and 5B respectively. It is very important to mention that those counts are subject to increase with time, hence enlarging the misleading ratios. Moreover, the misleading negative effect is much worse in real life context where billions of Reach counts may occur and vary from a community to another. In this context, additional research need to take place in order to identify the final destination of these tweets in order to take the needed actions for immediate remediation. 

\section{Impact Analysis of COVID-19 Related Tweets Initiated Per User Occupation/Specialty}
\label{credibility} 

Additional experiments were performed by considering the 83\% COVID related tweets in order to distinguish the identity of the tweeters initiating the unique tweets with COVID-19 context. We built an ontology of experts based on several dictionaries. The results of the ontology-based classification allowed us to study the profile of the 288K tweeters and identify 510 occupations belonging to the COVID tweet initiators. In this regard, we extracted very important insights about the credibility of tweets' initiators who might be eligible for broadcasting relevant messages in such a critical period. 

\begin{figure}
\centering
\includegraphics[scale=0.7]{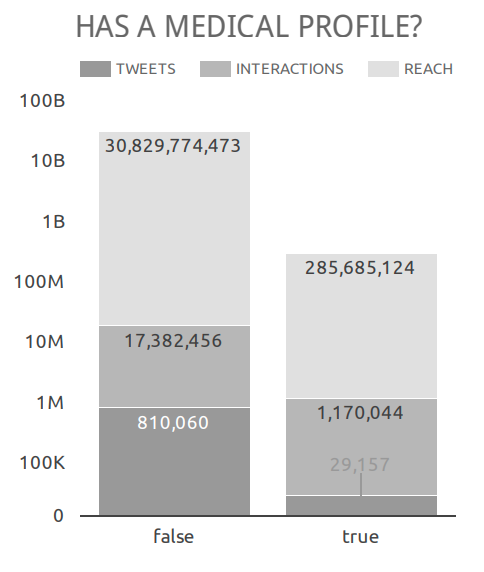}
\caption{Interactions and Reach Counts of the 3.5\% COVID Tweets Initiated by Medical Experts}
  \label{fig:fig6}
\end{figure}

\begin{figure}
\centering
\includegraphics[scale=0.7]{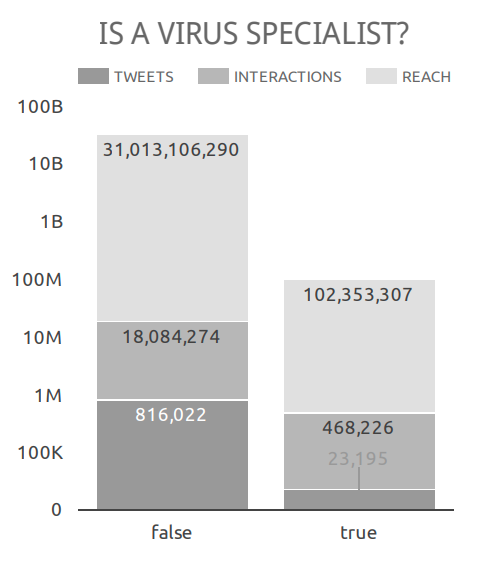}
\caption{Interactions and Reach Counts of the 2.8\% COVID Tweets Initiated by Virus Specialists}
  \label{fig:fig7}
\end{figure}

Among the 83.9\% of tweets, we first filtered the 839.2K Unique Tweets into Medical Profile and Non-Medical Profile categories based on the biographic information of each tweeter having at least one COVID context related tweet. Figure \ref{fig:fig6} aims at showing the participation of users that have medical backgrounds in the overall conversations in order to measure their effect based on their corresponding Interactions and Reach counts. It is clear that only 29.1K Tweets (i.e. 3.5\% of the COVID related tweets) were initiated by tweeters that have medical profiles, while the other 96.4\% of the tweets were initiated by tweeters that do not have medical profiles or expertise. Likewise, Figure \ref{fig:fig7} measures the different Interaction and Reach counts for the tweeters having virus specialty backgrounds. It also shows that only 2.8\% of the COVID related tweets were initiated by specialists, while the remaining 97.2\% were initiated by other tweeters' profiles. Usually, a specialty profile could be inherited from a medical profile, but not the opposite. We can depict from both Figures \ref{fig:fig6} and \ref{fig:fig7} that the total Interactions and Reach counts of tweets initiated by non-specialists tweeters are around 18M and 31B respectively, which reflect 38.6 and 303 times more than the tweets initiated by specialists in the field respectively. This might be very critical since it reflects the extent of the unintentional or intentional mislead ratios who may lead to potentially spreading unverified and non-credible medical information and guidelines for defeating COVID-19. 

\begin{figure*}
\centering
\begin{subfigure}[b]{0.32\textwidth}
        \includegraphics[width=\textwidth]{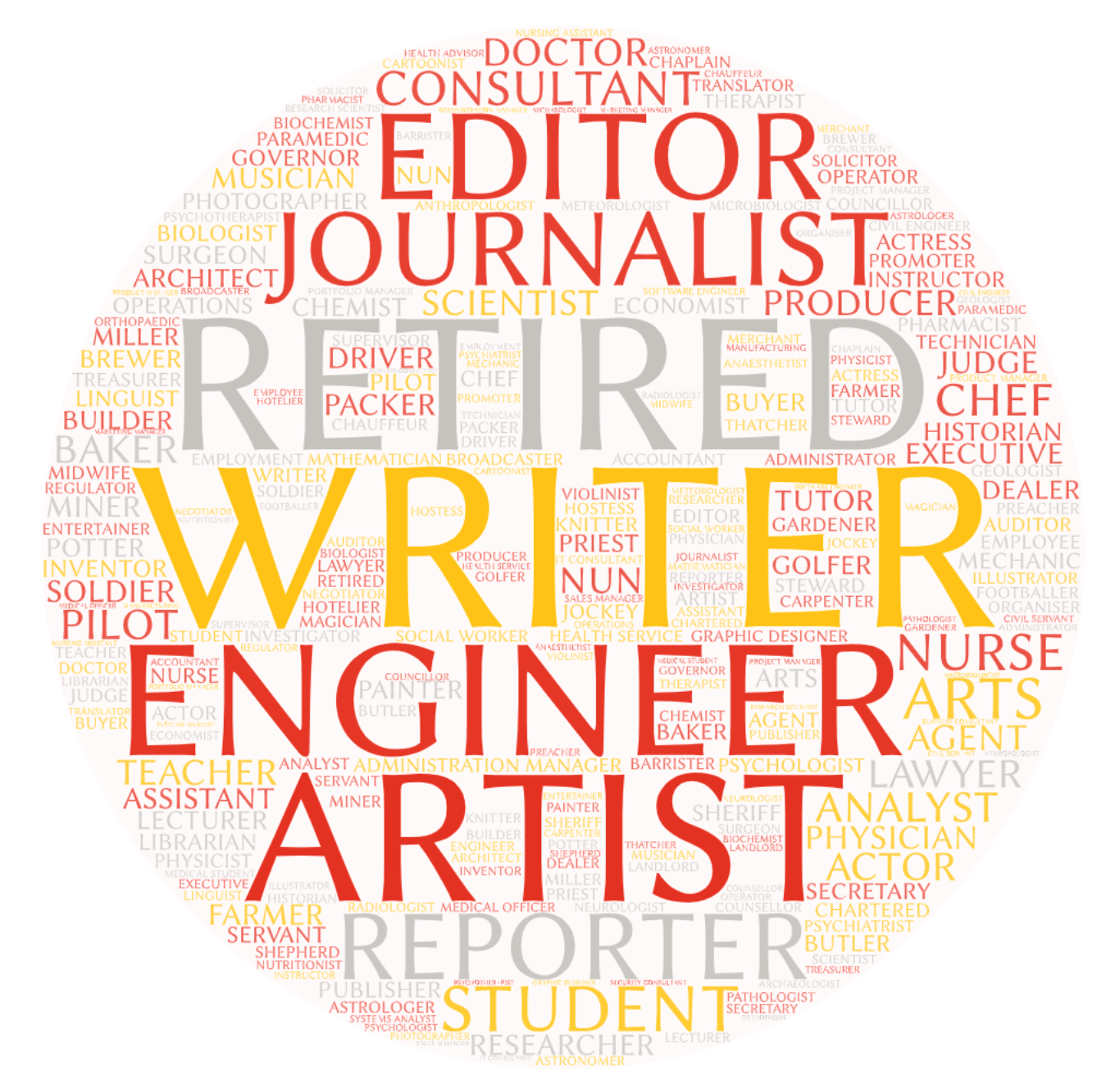}
         \caption{By Tweet Counts}
  %       \label{fig:}
    \end{subfigure}
    \begin{subfigure}[b]{0.32\textwidth}
        \includegraphics[width=\textwidth]{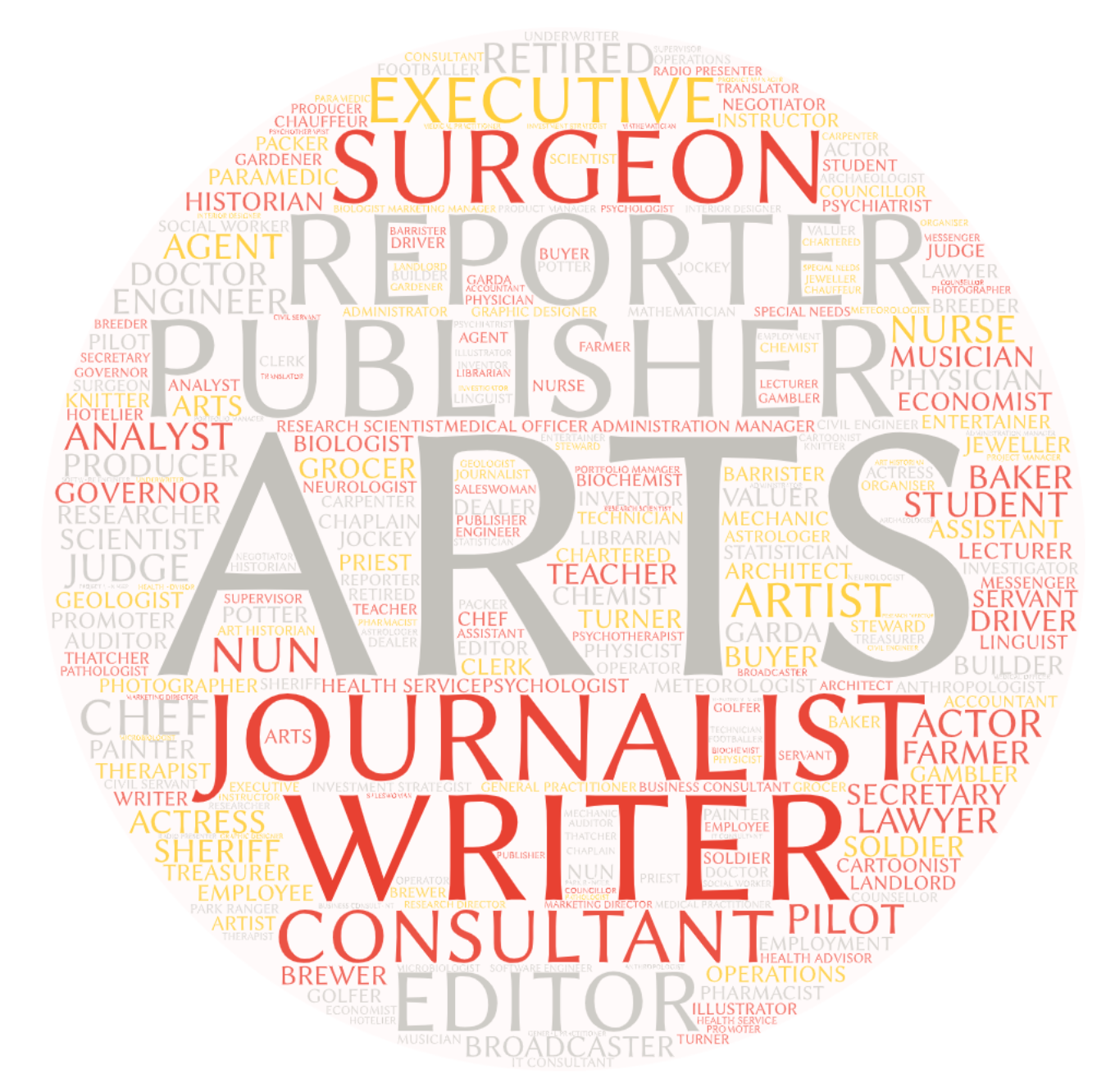}
        \caption{By Reach Counts}
     %     \label{fig:}
    \end{subfigure}
    \begin{subfigure}[b]{0.32\textwidth}
          \includegraphics[width=\textwidth]{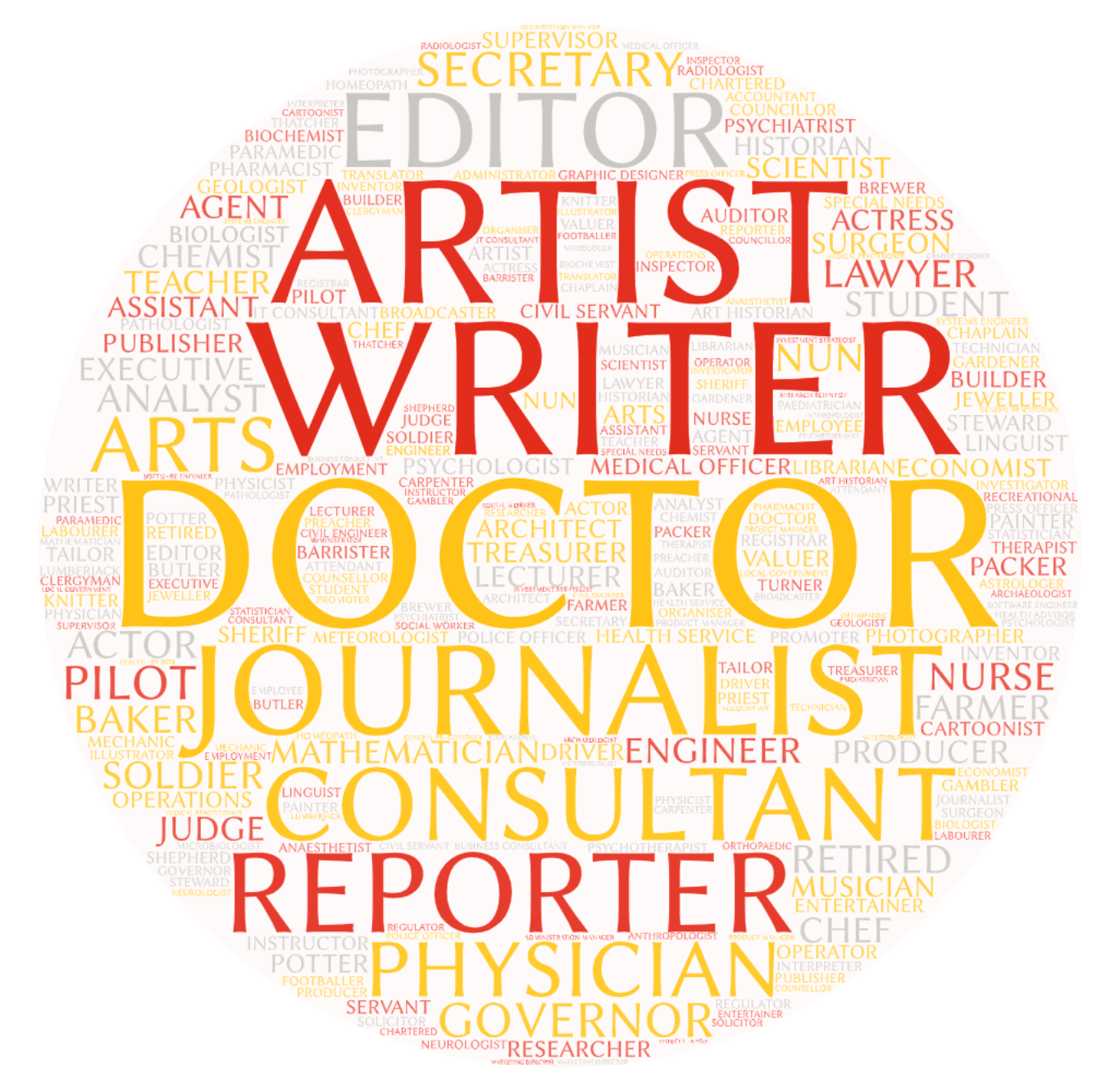}
        \caption{By Interactions Counts}
    %      \label{fig:}
    \end{subfigure}
    \caption{COVID Tweeters' Occupations/Specialities (Font Size Reflects the Count Value)}
    \label{fig:fig9}
\end{figure*}

\begin{figure}
\centering
\includegraphics[scale=0.62]{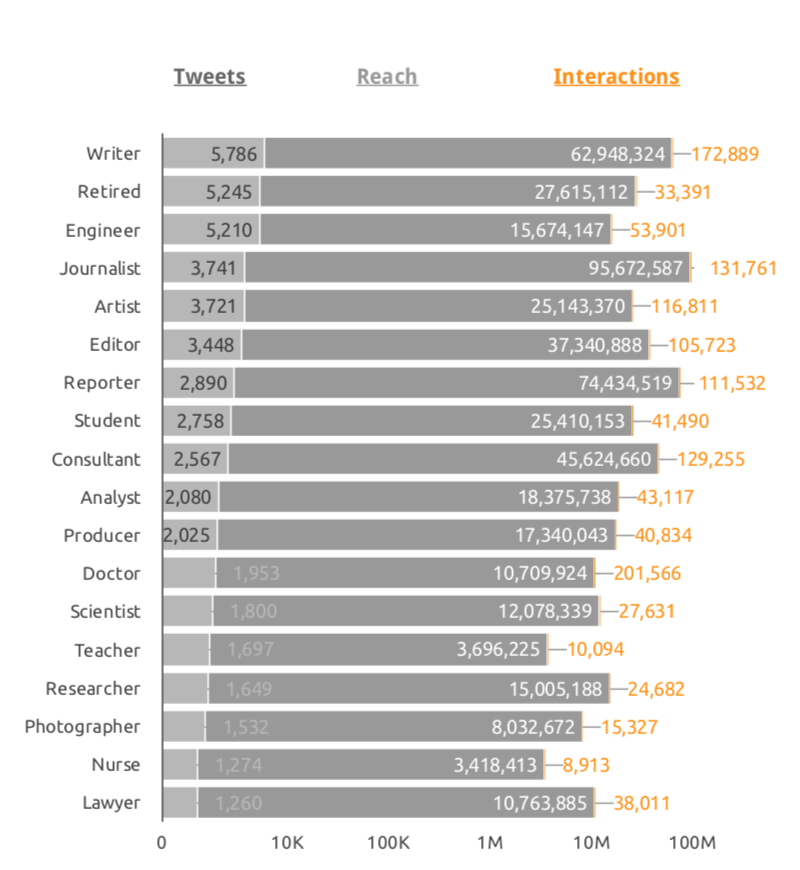}
\caption{COVID Tweet, Interactions and Reach Counts by Different Occupations/Specialities}
  \label{fig:fig8}
\end{figure}

The above implications should neither overshadow nor dominate the need for credible professional tweeters who should to contribute information that will raise awareness and defeat the virus. Governors, mayors, editors, writers and journalists are obvious examples of tweeters who should be on the list of occupations other than medically related who should be encouraged to interact and engage in such critical times. Indeed, the results of the COVID tweeters occupational analysis and classification based on their Tweet, Reach and Interactions counts support our hypothesis and findings. The list of credible tweeters could be expanded to include public figures such as actors and artists. Figure \ref{fig:fig9} presents three {\em wordles} that rearrange these occupations into a visual pattern broken down per Tweet, Reach and Interactions counts. The font size per occupation reflects its frequency while Figure \ref{fig:fig8} shows the top 18 occupations for the COVID tweeters' occupations initiating related unique tweets broken down per Tweet, Interactions and Reach. The main objective is to assess the impact of each group of tweeters and study their impact and influence rate in terms of Interactions and Reach. 
Clearly, both figures illustrate visually and numerically that the correlation between the number of Tweet, Interactions and Reach counts is not linear. In other words, the total Reach count of tweets initiated by the group of tweeters having {\em Arts} profiles and backgrounds are much higher than the total Reach of tweets initiated by the group of users having {\em Doctor} profiles and backgrounds, regardless of the number of uniquely initiated tweets by both groups. Furthermore, the correlation between the Tweet and Interactions counts is also not linear but logical. For instance, relevant occupations such as writers and journalists achieve high Interactions level, while non-related ones such as engineers and retired are getting low counts. Moreover, numerical results illustrate that context-related occupations such as doctors, writers, reporters, journalists, editors and governors do not even constitute 1\% of the total Reach counts, i.e. a total of around 300M out of 30B Reach counts. To further highlight the problem, these 1\% tweeters are supposed to be the only ones allowed to interact with people during such a critical situation. In this regard, two main implications can be reached from the presented results. First, accurate techniques are needed in order to verify the authenticity of the reported occupations based on historical and real-time means. Second, detection approaches need to be elaborated for identifying influencers relevant within specific contexts and situations.

\section{Implications \& Research Directions}
\label{directions} 

In this section, we provide various computing and non-computing implications, recommendation, and future research directions in relation to the aforementioned raised problems as inferred empirically and quantitatively: 

\begin{itemize}
\item An immediate ban should be placed on all the users, posts and tweets exploiting the COVID-19 context in order to mislead users and disseminate fake news. In this regard, various researchers tackled approached for detecting spams and misleading information in social networks based on users' meta-data, texts and contexts \cite{c31,c32,c33,c34,c35,c36,c37,c38,c39,c40,c41,c42,c43,c44,c45,c60}. However, these approaches did not consider critical and crisis times where high accuracy and time efficiency factors have major impact on overall solutions. Even major social network platforms have confirmed that applying current AI techniques without human interventions may lead to unfairness by wrongly banning valid accounts and interactions. Accordingly, additional research efforts have to investigate efficient and accurate human-less techniques and methodologies for better understanding the origin of misinformation while identifying both disruptive contexts and users.  

\item Although information broadcasts are not initiated by medical experts or officials, they may be at times essential and useful. Accordingly, allowing only communications by specific categories may be counterproductive as it could block legitimate and helpful information. In this regard, several approaches have addressed reputation and credibility based on user-centered and content-based analysis \cite{c46,c47,c48,c49,c50,c59,c52,c51,c53,c55,c56,c57,c58}. However, to the best of our knowledge, none of these approaches have classified and managed posts and accounts based on their verified roles, occupations and specialities. Consequently, mechanisms should be proposed in order to efficiently and accurately allow postings based on the aforementioned criteria, while at the same time considering credibility, historical engagement, insights and influence rate in related contexts and events. Moreover, there is a need at this time to develop systems that have efficient and highly accurate trust and credibility preserving models to be opportunistically adopted during crisis periods. Approaches relying on Blockchain and operating in highly distributed environments may be good options for potential efficient solutions regardless of their high cost.

\item Results show that the Reach level of professional COVID-19 context-relevant roles and occupations (e.g. doctors, editors, governors) is very low (i.e. only 1\% of total Reach). Accordingly, extensive effort should be put to elaborate methodologies and recommendation systems for efficiently recognizing credible and convincing influencers in specific events/locations/communities (e.g. based on profile, insights, historical engagement) for spreading the relevant and cited information provided by trusted scientists and experts at large scale, in the right place and to the right people. In this context, researchers may benefit from the rich literature that targets identifying influencers based on selected events in order to build relevant approaches \cite{c63,c64,c65,c66,c67,c68,c69,c70}.

\item Current raised {\em infodemic} shed the light on the urgent need to elaborate methodologies and techniques to be embedded in the social network platforms for systematically adopting emergency and crisis mode management strategies and responding to the situation dangers. This also includes developing code of conduct, standards and regulations to abide by during crisis periods, which may differ from the policies applied within regular terms. Although few approaches studied the role and reaction of social network platforms in response to previous natural disasters \cite{c77,c78}, the research field still lacks solid and sustainable methodologies to deal with epidemic and pandemic contexts, and prior, during and post crisis.

\item {\em Infodemic} made it difficult for people to find reliable resources for information. Accordingly, the UN is stepping up their communications efforts through global cooperation and viral acts of humanity. Although some are promoting the Chinese model of censored contagion, the solution is for health authorities, governments and social network stakeholders to formulate regular responses to the {\em infodemic} using a strategy of active engagement and communication with those who are spreading inaccurate stories in order to gain a deeper understanding of how {\em infodemic} spread. Governments should set-up official units mandated to combat the spread of inaccurate and unsubstantiated news. For example, the UK established a rapid response unit within the Cabinet Office. The Unit will work with social media firms in order to filter fake news and harmful content.

\item The most powerful solution to tackle this, or any future {\em infodemic}, lies with the consumers themselves. Taking personal responsibility of the role that each person plays when they receive, read, edit, comment and then forward a piece of information that originates on a social media platform is, arguably, the most impactful intervention to debunk the myths and falsehoods that are generated on an hourly basis. Targeted campaigns must be launched to educate anyone whose date of birth precedes the year 2000 to educate them on the social responsibility that they bear whenever they partake in perpetuating stories on {\tt Twitter} or any other platform.  

\end{itemize}

\section{Literature Review} 
\label{relatedwork}
In this section, we provide a literature review in relation to the aforementioned implications and proposed research directions, and which may form a solid ground for potential solutions. 

\subsection{Spam and Misleading Posts Detection}
Detecting spammers on social networks most often relies on analyzing the content of messages \cite{c6,c42, c45, c60, c44, c74}. However, most of the approaches extend their techniques by exploiting users’ profile and their relations\cite{c43}. Sedhai et al. \cite{c31} proposed a semi-supervised technique for spam detection, in which they proposed multiple detectors that investigate tweets' contents to classify maliciousness. Similarly, Alghamdi et al. \cite{c32} exploited a set of OSNs object and URL features for the same purpose. Such features include information related to user’s profile, and URL related features including hosts and domains. Similarly to the previous approach, Lee at al. \cite{c39} deployed a real-time malicious URL detector by exploiting URL redundancy driven by the limitation posed on the attackers' resources. Guille et al. \cite{c38} proposed another approach that takes advantage of the URL used by the users in their tweets to spot malicious intents. A multi-feature analysis like unique mentions, trends, hyperlinks, and tweets ratio has been employed by Amleshwaram et al. \cite{c33} to distinguish spam accounts in a supervised manner. Moreover, Benevenuto et al. in \cite{c34} aimed to classify users between promoters, spammer and legitimated from their videos. By manually selecting different users and learning their behaviors, authors were able to employ a supervised machine learning technique capable of classifying malicious users with a relatively small margin error. Kuhn et al. \cite{c35} described spamming strategies techniques of more than 570 million tweets. SHEN et al. \cite{c41} deployed another approach that depends on the tweets contents to extract users’ behaviors and supply them to a supervised classifier. However, supervised and semi-supervised techniques cannot classify data by discovering features on their own, which requires manual classification in the initial stages. Such involvement requires the intervention of human in which by its nature prone to errors, thus reducing the accuracy of the results. In case of online social networks, classification of diverse and large amount of data has been proven difficult. 

\subsection{User-Centered \& Content-Based Reputation \& Credibility Analysis}
Despite the work on detecting spammers in social networks, other approaches took advantage of the abundance number of information for ranking users based on their influence rate. Such techniques stem from the need to rank the relevance of the users and their tweets, and thus two main categories of solutions exist to address the issue in dispute. The first set of approaches focused on the content to assign reputation using machine learning techniques \cite{c61, c62, c63}, while the second set relied on the user and its relation described as nodes in a graph model \cite{c61, c64, c47}. Moreover, there are other solutions that depend on both methods to achieve better accuracy. In the following, we overview the main approaches belonging to these categories.

Jain et al. \cite{c46} took advantage of the capabilities of graph theories and related algorithms to calculate a score for each user based on their centralities. Such scores are later used to identify universal leaders' opinions. Riyantoa et al. \cite{c47} provided an in-depth analysis on how social distancing and environment can affect trust and trustworthiness between users. Mohammadinejad et al. \cite{c50} presented a framework that takes advantage of the consensus opinion within social network relations to infer scores such as user’s personality to derive the most influential users in the network. Zhang et al. \cite{c51} benefited from the relations through social network messages and contact frequency to learn the user’s behavior, thus providing a credibility score that describes the risk levels of users' interactive messages. Wang et al. \cite{c59} provided an empirical analysis on the information credibility and provided a credibility assessment framework. They also emphasized the value of users' credibility in relation to the credibility of the information. Tsikerdekis et al. \cite{c52} drew the attention towards recent adversaries related to social network including identity deception and multiple account creation, and employed a behavioral framework to detect such actions. 

Ahmad et al. \cite{c58} presented a survey on different approaches used for the detection of rumors on social networks. Curiskis et al \cite{c53} provided a comparison of different document clustering techniques that are mostly used on OSNs and supplied by multiple features. Moreover, they also provided several evaluation measures to assess their accuracy. Buzz et al. \cite{c53} focused on the content in different languages such case “Arabic” in order to produce a framework that is able to distinguish fake news by allotting a score for each content through sentiment analysis with the help of different classification algorithms. Alrubaian et al. \cite{c56} proposed a system with multiple components that work in conjunction to deduce the credibility of users and their related tweets to restrain the spread of fake and malicious news. 

\subsection{Influence Ranking in Social Networks} 
Users' influence rating and ranking have become one of the most important topics when analyzing social networks, especially in microblogs like {\tt Twitter}. Authors in \cite{c62, b9, b11, b12 , c68, c69} explored that user meta data like follower count, tweets count, following count and tweets meta data like retweet count and favorite count are enough to calculate the user influence ratio. On the other hand, authors in \cite{b13} analyzed the relationships between users in order to rank them by their influence relationships. \cite{b14} analyzed the user's social activity during a specific event. Anger, Isabel \&amp and Kittl, Christian \cite{b15} determined a grounded approach to measure the individual's influence or potential social networking ratio (SNP) using users and tweets metadata to find the top 10 {\tt Twitter} users in Austria. Bakshy, Eytan et. al \cite{b16} calculated the user influence rate per event using diffusion trees and cascading methods by selecting only events that have URLs. Then, they applied diffusion algorithms on the shared URLs to measure the reach of the initial tweets. M. Anjaria and R. M. R. Guddeti \cite{b17} used NLTK sentiment analysis and Incremental Learning algorithms to predict the presidential elections in the US. Moreover, C. B. Schenk and D. C. Sicker \cite{b18} categorized influencers into four influence groups using a bagging classification algorithm by studying users static and dynamic influence features and comparing them over time. 

In \cite{b19}, Y. Mei, Y. Zhong and J. Yang approached an entropy weighting algorithm based on eight data points per each user to find their influence ratios. They added the features of new followers and new mentions to measure users' popularity ratios in order to sort a list of the top hundred users in Australia by their influence rates. Riquelmea et al. \cite{ c63} proposed two linear threshold centrality based approaches to measure the rank of the users and the propagation rate of their contents in the network. Similarly, Li et al. \cite{c64} presented an eigenvector centrality based approach to measure the influence rate. Lahuerta-Otero et al. \cite{ c65} presented a brief analysis of the behavior of special kind of tweeter users, and evaluated their influence ratio through different data mining techniques. Through their analysis, they were able to spot different techniques to increase users’ influence. Sharma et al. \cite{c66} proposed a novel approach to elect influential users by calculating the influence rate through their tweet and trend scores. Huynh et al. \cite{c70} focused on the relation between the tag used in the tweets to calculate the influence rate and the speed of their propagation.

\section{Conclusion}
\label{conclusion} 

This paper investigated the COVID-19 {\em infodemic} negative impact on the major efforts to defeat the pandemic through a novel large-scale Twitter-based study, which provided quantitative assessment using real-life experiments reflecting the actual environments. The empirical analysis of 1 million COVID-19-related tweets belonging to 288K unique users illustrated the severe impact of misleading people and spreading unreliable information. Inferred insights showed that (1) the potential reachability of the 16.1\% tweets that misled users by redirecting them to out of scope and/or malicious content is 5.6 billion, and (2) a minimum of 93.7\% of the remaining within-context 83.9\% tweets (i.e. with around 17M Interactions and 30B Reach counts) were initiated by users with non-reliable medical and/or relevant speciality profiles, and consequently might be disseminating misleading non-credible medical information. Moreover, different insights highlighted the low reachability (i.e. 1\% of the total Reach counts, which is equivalent 300M out of 30B) of the unique users with key context-relevant specialties and occupations such as doctor, writer, reporter, journalist, editor and governor. The results shed the light on the importance of identifying non-medical key influencers for assisting in spreading legitimate information relevant in such situations. Finally, the paper elaborated on few computing and non-computing implications as well as future research directions to highlight the potential solutions and future work in such a promising field.

\ifCLASSOPTIONcaptionsoff
  \newpage
\fi

\bibliographystyle{IEEEtran}
\bibliography{References}
\textbf{\\Biographies}
\noindent {\\\\\bf \small Azzam Mourad} \small is currently an Associate Professor of Computer Science at the Lebanese American University and also an Affiliate Associate Professor at the Software Engineering and IT Department, Ecole de Technologie Superieure (ETS), Montreal, Canada. He has served/serves as Associate Editor for the IEEE Transaction on Network and Service Management, IEEE Network, IEEE Open Journal of the Communications Society, IET Quantum Communication, and IEEE Communications Letter. He has also served/serves as General Chair of IWCMC2020, General Co-Chair of WiMob2016, and  Track Chair, TPC member, and Reviewer of several prestigious journals and conferences. He is an IEEE senior member. 

\noindent {\\\\\bf \small Ali Srour} \small is currently pursuing his Masters in Computer Science at the Lebanese American University. He is a Data Scientist and an AI consultant. His research interests are Social Network Analysis, Data Science and Artificial Intelligence Innovations.

\noindent {\\\\\bf \small Haidar Harmanai} \small received his BS, M.S, and Ph.D. in Computer Engineering from the Department of Electrical
Engineering and Computer Science at Case Western
Reserve University, Cleveland, Ohio, in 1989, 1991, and 1994 respectively. He is currently a Professor of Computer Science at the Lebanese American University.  He serves on the steering committee of the IEEE NEWCAS conference and the IEEE ICECS
conference. He has also served on the program committees of various international conferences. His research interests include electronic design automation, high-level synthesis, design for testability, and parallel programming.  He is a senior member of IEEE and a senior member of ACM. 

\noindent {\\\\\bf \small Cathia Jenainati} \small is currently a Professor of English Literature and the Dean of the School of Arts and Sciences at the Lebanese American University. Dr. Jenainati has previously served as the founding head of the School for Cross-Faculty Studies at Warwick University, UK. She serves as the Associate Editor for the Journal of Coaching Practice, Chair of the International Advisory Board of Amsterdam University College, Educational Coach of the Growth Coaching International (Australia – UK – USA), and Founding Fellow of the Warwick Higher Education Academy (UK). Her research focuses on women’s activism, oral history, the global sustainable development agenda, and the history of education missions in the Middle East. In addition, she has been recognized as a leader in innovative pedagogies especially around Liberal Education, and received several research grants in the field. 

\noindent {\\\\\bf \small Mohamad Arafeh} \small is currently a research assistant at the Lebanese American University. He holds an M.Sc. in Business computer and information systems from the Lebanese University. His research interests are crowdsensing, social network analysis and blockchain technology.

\end{document}